\documentstyle[preprint,eqsecnum,aps,epsf]{revtex}	% preprint format

\newif\iftightenlines\tightenlinesfalse
\tightenlines\tightenlinestrue

\begin{document}
%
%%%%%%%%%%%%%%%%%%%%%%%%%%%%%%%%%%%%%%%%%%%%%%%%%%%%%%%%%%%%%%%%%%%%%%%%%%%%%%%
\def\mol{Mol}
\def\eslt{E\llap/_T}
\def\esl{E\llap/}
\def\msl{m\llap/}
\def\to{\rightarrow}
\def\te{\tilde e}
\def\tmu{\tilde\mu}
\def\ttau{\tilde\tau}
\def\tl{\tilde\ell}
\def\ttau{\tilde \tau}
\def\tg{\tilde g}
\def\tnu{\tilde\nu}
\def\tell{\tilde\ell}
\def\tq{\tilde q}
\def\tb{\tilde b}
\def\tst{\tilde t}
\def\tt{\tilde t}
\def\tw{\widetilde W}
\def\tz{\widetilde Z}

\hyphenation{mssm}
%\def\ds{\displaystyle}
%\def\ts{${\strut\atop\strut}$}
%
%%%%%%%%%%%%%%%%%%%% TITLE PAGE %%%%%%%%%%%%%%%%%%%%%%%%%%%%%%%%%%%%%%%%%%%%%%
\preprint{\vbox{\baselineskip=14pt%
   \rightline{FSU-HEP-970605}\break 
%   \rightline{Preliminary draft}\break 
}}
\title{Neutralino Dark Matter in Minimal Supergravity:\\
Direct Detection vs. Collider Searches}
\author{Howard Baer and Michal Brhlik}
\address{
Department of Physics,
Florida State University,
Tallahassee, FL 32306 USA
}
\date{\today}
\maketitle
\begin{abstract}

We calculate expected event rates for direct detection of relic neutralinos
as a function of parameter space of the minimal supergravity model.
Numerical results are presented for the specific case of a $^{73}$Ge
detector.
We find significant detection rates ($R> 0.01$ events/kg/day) 
in regions of parameter space most favored
by constraints from $B\to X_s\gamma$ and the cosmological relic
density of neutralinos. 
The detection rates are especially large in regions
of large $\tan\beta$, where many conventional signals for supersymmetry
at collider experiments are difficult to detect. 
If the parameter $\tan\beta$ is large, then there
is a significant probability that the first direct evidence for
supersymmetry could come from direct detection experiments, 
rather than from collider searches for sparticles.
\end{abstract}

\medskip
%{\tt$\backslash$\string pacs\{\}}

%\narrowtext

%%%%%%%%%%%%%%%%%% MAIN TEXT %%%%%%%%%%%%%%%%%%%%%%%%%%%%%%%%%%%%%%%%%%%%%%%

\section{Introduction}

Luminous matter comprises less than $\Omega_{lum} =\rho /\rho_c \sim 0.01$ 
of the matter density of the universe\cite{reviews,jung}, where 
$\rho_c={3H_0^2\over {8\pi G_N}}$ is the critical closure density of the 
universe, and $H=100h$ km/sec/Mpc is the scaled Hubble constant, 
with $0.5\alt h \alt 0.8$. In contrast, most inflationary cosmological 
models require $\rho =\rho_c$, {\it i.e.} a flat universe. The difference
in these matter density values can be reconciled by hypothesizing the existence
of dark (non-shining) matter (DM) in the universe.
Experimental evidence for galactic dark matter comes from 
the enclosed mass versus velocity plots measured for clouds of neutral
hydrogen rotating about
galactic centers (galactic rotation curves), which imply
$\Omega\ge 0.03- 0.1$. An understanding of galactic clustering and 
galactic flows points towards even larger values of $\Omega \sim 0.2-1$, 
possibly in accord with inflationary cosmology. 
Calculations of Big Bang nucleosynthesis
can only allow the baryonic contribution to the matter density of the universe 
to be $\Omega_{baryonic}\sim 0.01-0.1$, so that if $\rho =\rho_c$, 
the bulk of dark 
matter in the universe must be non-baryonic. Some candidates for non-baryonic
dark matter from particle physics include neutrinos with eV scale masses
(hot dark matter: HDM) and
WIMPs (weakly interacting massive particles), such as axions or the
lightest neutralino in supersymmetric (SUSY) models. Some compelling 
models of structure formation in the universe which take into account
COBE measurements of the anisotropy in the 
cosmic microwave background actually prefer a ``mixed dark matter'' (MDM)
universe, with $\sim 10\%$ baryonic matter, $\sim 30\%$ HDM,
and $\sim 60\%$ cold dark matter (CDM, comprised of WIMPs).

In this paper, we focus attention on the lightest SUSY particle, 
or LSP--- usually the lightest neutralino--- from supersymmetric models. 
At very early times after the Big Bang, neutralinos would have existed in 
thermal equilibrium with the primordial cosmic soup of particles and
radiation. As the universe expanded and cooled, temperatures dropped so 
low that neutralinos could no longer be produced, although they could still
annihilate away. Ultimately, the expansion rate of the universe exceeded
the annihilation rate so that a relic density of neutralinos would 
be locked in. Numerous estimates of the neutralino relic density 
$\Omega h^2$ as a function of SUSY model parameter space have been 
made\cite{relic,bb}. 
Models that predict $\Omega h^2<0.025$ cannot even account for the 
dark matter needed for galactic rotation curves, while values of
$\Omega h^2>1$ would yield a universe younger than 10 billion years old, 
in contradiction at least with the ages of the oldest stars found in
globular clusters.

A consequence of the SUSY dark matter hypothesis is that a non-relativistic 
gas of neutralinos fills all space. To test this hypothesis,
a number of direct detection experiments have been built or are 
under construction\cite{detect}. 
The general idea behind these experiments is that
relic neutralinos (or other possible WIMPs) could scatter off the nuclei in
some material, depositing typically tens of keV of energy. 
Some examples of how the thermal energy 
could be detected include: {\it i.}) via changes in 
resistance due to a slight temperature increase (bolometry), 
{\it ii.}) via a magnetic flux change due to a superconducting granule 
phase transition, or {\it iii.}) via 
ionization. The technical challenge is to build detectors that could pick
out the relatively rare, low energy neutralino scattering events 
from backgrounds mainly due to cosmic rays and radioactivity in 
surrounding matter.
Future detectors are aiming to reach a sensitivity 
of 0.1-0.01 events/kg/day. In this way, the first evidence for SUSY 
might come from direct neutralino detection rather than from 
accelerator experiments.

Since the pioneering paper by Goodman and Witten\cite{gw},
calculations of neutralino-nucleus scattering have seen continual 
improvements\cite{others,dn,jung}.
The first step involved in a neutralino-nucleus scattering calculation is to
calculate the effective neutralino-quark and neutralino-gluon interactions.
The neutralino-quark axial vector interaction leads in the non-relativistic
limit to a neutralino-nucleon spin-spin interaction, which involves the 
measured quark spin content of the nucleon. To obtain the neutralino-nucleus
scattering cross section, a convolution with nuclear spin form factors 
must be made. The neutralino-quark and neutralino-gluon interactions 
(via loop diagrams) can also resolve into scalar and tensor components. 
These interactions can be converted into an effective scalar 
neutralino-nucleon interaction involving quark and gluon parton distribution
functions. A neutralino-nucleus scattering cross section can be obtained
by convoluting with suitable scalar nuclear form factors. The final neutralino 
detection rate is obtained by multiplying by the {\it local} neutralino
density (estimates are obtained from galaxy formation modeling), and
appropriate functions involving the velocity distribution of relic
neutralinos and the earth's velocity around the galactic center.

In this paper, we calculate event rates for 
direct detection of relic 
neutralinos left over from the Big Bang. For illustration, we present detailed
calculations for a $^{73}$Ge detector; a $^{73}$Ge detector has 
a sizable nuclear spin content $J={9\over 2}$, so that it would be 
sensitive to both spin and scalar neutralino-nucleus interactions\cite{dep}. 
Our goal is to compare the reach for
supersymmetry by direct detection experiments with constraints on model
parameter space from relic density calculations, collider search limits, and
limits from $B\to X_s\gamma$ measurements. In addition, we will compare
the reach of direct detection experiments with the previously calculated
reach of collider facilities such as LEP2, the 
Fermilab Tevatron Main Injector (MI) upgrade and 
the CERN LHC $pp$ collider.

We work within the framework of the paradigm minimal supergravity (mSUGRA)
model\cite{sugra}. 
This model assumes the minimal supersymmetric standard model, or MSSM, 
is valid at all energy scales from $M_{weak}$ up to 
$M_{GUT}\simeq 2\times 10^{16}$ GeV. The mSUGRA model could arise as the low 
energy limit of a supergravity theory, where supersymmetry is broken in the
hidden sector of the model at energy scale $M\sim 10^{10}$ GeV. 
Supersymmetry breaking is communicated to the observable sector via 
gravitational interactions, leading to soft SUSY breaking mass terms
of order the electroweak scale, $\tilde{m}\sim 100-1000$ GeV. 
At the GUT scale (with the added assumption of an approximate 
global $U(n)$ symmetry for
the mSUGRA Lagrangian),
this leads to a common mass for all scalars $m_0$ and a common trilinear
coupling $A_0$. Motivated by the apparent unification of gauge coupling 
constants, it is also assumed that all gaugino masses are unified to
$m_{1/2}$ at $M_{GUT}$. The weak scale sparticle spectrum is derived from 
renormalization group (RG) running of the SUSY soft breaking parameters.
Requiring radiative electroweak symmetry breaking allows the determination 
of the superpotential Higgsino mass squared $\mu^2$, and allows the expression of
the soft SUSY breaking bilinear term $B$ in terms of $\tan\beta$, the ratio of
vev's of the two Higgs fields. Thus, all sparticle masses and couplings are 
derived in terms of the parameter set 
\begin{eqnarray*}
m_0,\ m_{1/2},\ A_0,\ \tan\beta ,\ {\rm and}\ sign(\mu ).
\end{eqnarray*}
We adopt the mSUGRA spectrum calculation encoded into the event generator 
ISAJET 7.29\cite{isajet}.

In Sec. 2, we present details of our calculation for neutralino
scattering events off a $^{73}$Ge detector. In Sec. 3 we present 
numerical results for event rates in the parameter space of the mSUGRA
model, and compare to neutralino relic density contours. 
In Sec. 4, we compare our results to $B\to X_s\gamma$ constraints, and to 
collider search constraints. 
Finally, we compare the 
search capability of a $^{73}$Ge dark matter detector to the reach of
various future collider experiments for mSUGRA. We find usually 
that a DM detector attaining a sensitivity of 
0.01 events/kg/day will have a greater reach into mSUGRA parameter space
than either the LEP2 or Tevatron upgrades, via their searches for 
sparticles. However, such dark matter detectors can only probe a fraction
of the parameter space that gives rise to a reasonable relic density.
A complete exploration of the cosmologically interesting mSUGRA 
parameter space will have to await the CERN LHC $pp$ collider, 
or an $e^+e^-$ or $\mu^+\mu^-$ collider operating at $\sqrt{s}\sim 1$ TeV. 
In Sec. 5, we give a summary and some conclusions.

\section{Calculational details}

\subsection{Dark Matter Detection: Theory}

The effective elastic scattering Lagrangian can generally be divided into two
parts:
\begin{eqnarray}
{\cal L}^{eff}_{elastic}={\cal L}^{eff}_{scalar}+{\cal L}^{eff}_{spin}.
\end{eqnarray}
We examine first the scalar Lagrangian, which receives contributions from
neutralino-quark interaction via squarks and Higgs bosons exchange, and from
neutralino-gluon interactions at one-loop level involving quarks, squarks and
Higgs bosons in the loop diagrams. On the parton level it is expressed 
at scale $Q$ (typically $\sim m_{\widetilde Z_1}$) as \cite{dn} 
\begin{eqnarray}
{\cal L}^{eff}_{scalar}&=&f_q \bar{\tz_1}\tz_1 \bar{q}q
+g_q \left[-2i \bar{\tz_1}\gamma_{\mu}\partial_{\nu}\tz_1 {\cal Q}^{(2)\mu\nu}
-\frac{1}{2}m_q m_{\widetilde Z_1}\bar{\tz_1}\tz_1 \bar{q}q \right] \cr
& &+\alpha_S \left[-(B_{1D}+B_{1S})\bar{\tz_1}\partial_{\mu}\partial_{\nu}
\tz_1  
+B_{2S} \bar{\tz_1}(i\gamma_{\mu}\partial_{\nu}+i\gamma_{\nu}\partial_{\mu})
\tz_1
\right] {\cal G}^{(2)\mu\nu}\cr
& &+\alpha_S b \bar{\tz_1}\tz_1 F^{a}_{\mu\nu}F^{a\mu\nu}.
\end{eqnarray}                                 
Here 
\begin{eqnarray*}
{\cal Q}^{(2)}_{\mu\nu}&=&\frac{i}{2}(\bar{q}\gamma_{\mu}\partial_{\nu} q+
\bar{q}\gamma_{\nu}\partial_{\mu}
q-\frac{1}{2}g_{\mu\nu}\bar{q}\partial\llap/q ) \cr
{\cal G}^{(2)}_{\mu\nu}&=&F^{a}_{\mu\rho}F^{a\rho}_{\ \ \nu}
+\frac{1}{4}g_{\mu\nu}F^{a}_{\rho\sigma}F^{a\rho\sigma}
\end{eqnarray*}
are traceless twist-2 quark and gluon operators, and for the sake of brevity
the effective couplings are given in the Appendix. Using nucleonic matrix
elements
\begin{eqnarray*}
\langle N|{\cal Q}^{(2)}_{\mu\nu}(q^2\rightarrow 0)|N\rangle &=&\frac{1}{m_N}(p_{\mu} p_{\nu} 
-\frac{1}{4}m_N^2 g_{\mu\nu})\int_{0}^{1} x [q_N (x,Q)+\bar{q}_N(x,Q)]\, dx \cr
\langle N|{\cal G}^{(2)}_{\mu\nu}(q^2\rightarrow 0)|N\rangle &=&\frac{1}{m_N}(p_{\mu}p_{\nu} 
-\frac{1}{4}m_N^2 g_{\mu\nu})\int_{0}^{1} x g_N (x,Q) \, dx ,
\end{eqnarray*}     
(here, $q^2$ is the momentum transfer squared, and $p_\mu$ is the nucleon 
four-momentum)
and introducing the parton distribution functions $q_N$, $\bar{q}_N$ and 
$g_N$, and using
\begin{eqnarray*}
\langle N|F^{a}_{\rho\sigma}F^{a\rho\sigma}(q^2\rightarrow 0)|N\rangle &=&\frac{8\pi}
{9\alpha_S} m_N f_{TG}^{(N)},\cr
\langle N|m_q \bar{q}q|N\rangle &=& m_N f_{Tq}^{(N)},
\end{eqnarray*}
it is possible to convert Lagrangian (2.15) into an effective
neutralino-nucleon Lagrangian
\begin{eqnarray}
{\cal L}^{eff}_{scalar}&=&f_p \bar{\tz_1}\tz_1 \bar{\Psi}_p \Psi_p+
f_n \bar{\tz_1}\tz_1 \bar{\Psi}_n \Psi_n.
\end{eqnarray}   
Evaluation of $f_N$ for $N=p,n$ yields
\begin{eqnarray}
\frac{f_N}{m_N}&=&\sum_{\scriptstyle u,d,s} \frac{f_{Tq}^{(N)}}{m_q} 
[f_q^{(\tilde q)}+f_q^{(H)}-\frac{1}{2} m_q m_{\widetilde Z_1} g_q]
+\frac{2}{27}f_{TG}^{(N)} \sum_{\scriptstyle c,b,t} \frac{f_{q}^{(H)}}{m_q}\cr
& &-\frac{3}{2} m_{\widetilde Z_1}\sum_{\scriptstyle u,d,s,\atop c,b } 
g_q(Q) (q_N (2,Q)+\bar{q}_N (2,Q)) -\frac{8\pi}{9} b f_{TG}^{(N)} \cr
& &+\frac{3}{2}\alpha_S m_{\widetilde Z_1} [B_{2S}+\frac{1}{2}
m_{\widetilde Z_1} (B_{1D}+B_{1S})] g_N (2,Q),
\end{eqnarray}  
where the various coupling constants are given in the Appendix.
Here we have used the general definition of an $n$-$th$ integral moment
$f(n,Q)=\int_{0}^{1} x^{n-1} f(x,Q)\, dx$ and applied it to the parton 
distribution functions. 
The differential cross section for a neutralino
scattering off a nucleus $X_{Z}^{A}$ with mass $m_A$ is then expressed as 
\begin{eqnarray}
\frac{d\sigma^{scalar}}{d|\vec{q}|^2}=\frac{1}{\pi v^2}[Z f_p +(A-Z) f_n]^2 
F^2 (Q_r),
\end{eqnarray}                                 
where $\vec{q}=\frac{m_A m_{\widetilde Z_1}}{m_A+m_{\widetilde Z_1}}\vec{v}$ is
the transfered momentum, $Q_r=\frac{|\vec{q}|^2}{2m_A}$ and $F^2(Q_r)$ is the
scalar nuclear form factor.

Interaction between the neutralino and quark spins is described by a 
parton-level Lagrangian \cite{jung}
\begin{eqnarray}
{\cal L}^{eff}_{spin}&=&d_q \bar{\tz_1}\gamma^{\mu} \gamma_5\tz_1 
\bar{q}\gamma_{\mu} \gamma_5 q
\end{eqnarray}                                 
which translates with the help of nucleonic spin matrix elements
\begin{eqnarray*}
<N|\bar{q}\gamma_{\mu} \gamma_5 q|N>=2 s_{\mu} \Delta q^{(N)} 
\end{eqnarray*}     
into 
\begin{eqnarray}
{\cal L}^{eff}_{spin}&=&2\sqrt{2}\left( a_p \bar{\tz_1}\gamma^{\mu} \gamma_5
\tz_1 
\bar{\Psi}_p s_{\mu}\Psi_p+
a_n \bar{\tz_1}\gamma^{\mu} \gamma_5\tz_1 \bar{\Psi}_n s_{\mu} \Psi_n\right) ,
\end{eqnarray}   
explicitly involving the nucleon spin vectors $s_{\mu}$. Coefficients 
\begin{eqnarray}
a_p=\frac{1}{\sqrt{2}} \sum_{\scriptstyle u,d,s} d_q \Delta q^{(p)},
&\ \ \  &
a_n=\frac{1}{\sqrt{2}} \sum_{\scriptstyle u,d,s} d_q \Delta q^{(n)}
\end{eqnarray}  
depend on experimental values of $\Delta q^{(N)}$, which are affected by
significant uncertainties which lead to variations in the cross section.
More details on the couplings are again given in the Appendix. 
For a nucleus with total
angular momentum $J$, the spin interaction differential cross section 
takes the form   
\begin{eqnarray}
\frac{d\sigma^{spin}}{d|\vec{q}|^2}=\frac{8}{\pi v^2}\Lambda^2 J (J+1) 
\frac{S(|\vec{q}|)}{S(0)},
\end{eqnarray}  
where $\frac{S(|\vec{q}|)}{S(0)}$ is the nuclear spin form factor normalized to
1 for pointlike particles, and $\Lambda=\frac{1}{J} [a_p \langle S_p\rangle 
+a_n \langle S_n\rangle ]$.
The quantities $\langle S_p\rangle$ and $\langle S_n\rangle$ represent the expectation value of the
proton (neutron) group spin content in the nucleus.

Putting both scalar and spin interaction contributions together and convoluting
them with the local neutralino flux (which depends on the {\it local} 
relic density $\rho_{\widetilde Z_1}$), the 
differential detection rate is calculated to be
\begin{eqnarray}
\frac{dR}{dQ_r}&=&\frac{4}{\sqrt{\pi^3}}\frac{\rho_{\widetilde Z_1}}
{m_{\widetilde Z_1} v_0} T(Q_r)
\biggl \{ [Z f_p +(A-Z) f_n]^2 F^2 (Q_r)\cr
& &\ \ \ \ \ \ + 8\Lambda^2 J (J+1) \frac{S(|\vec{q}|)}{S(0)} \biggr\},
\end{eqnarray}    
where $v_0\sim 220\,\rm km/s$ is the circular speed of the Sun around the
center of our galaxy and 
\begin{eqnarray}
T(Q_r)=\frac{\sqrt{\pi}v_0}{2}\int_{v_{min}}^{\infty}
\frac{f_{\widetilde Z_1}}{v}\, dv 
\end{eqnarray}   
integrates over the neutralino velocity distribution.

\subsection{Detection Rates in Germanium Detectors}

In order to obtain a quantitative estimate of neutralino detection rates in the 
mSUGRA framework which could be compared to phenomenological constraints and
collider reaches, we have evaluated the rate for the case of a $\rm ^{73}$Ge
detector.
In our calculations, we have taken the neutralino density
to be the {\it local} relic density consistent with galactic
formation models including varying baryonic mass fraction in the galaxy
\cite{flor}.
In the simplest Gaussian model taking into account
the motion of the Sun and Earth, the integrated velocity distribution can be
written as \cite{jung}
\begin{eqnarray}
T(Q_r)=\frac{\sqrt{\pi}v_0}{4v_e}
\left[ Erf(\frac{v_{min}+v_e}{v_0})-Erf(\frac{v_{min}-v_e}{v_0})\right]
\end{eqnarray}     
with $v_{min}=\sqrt{\frac{Q_r (m_{\widetilde Z_1}+m_A)^2}{2 m^2_{\widetilde Z_1}
m_A}}$, and where the Earth velocity $v_e$ is given by
\begin{eqnarray}
v_e = v_0\left[ 1.05+0.07\cos (\frac{2\pi (t-t_p)}{1\ yr} )\right],
\end{eqnarray}
with $t_p\simeq$ June 2.
The general nuclear properties needed are the $\rm ^{73}Ge$ mass, 
$m_{Ge}=67.93$ GeV, and its total spin $J=\frac{9}{2}$. 

To determine the scalar contribution, it is necessary to compute the
parton distribution integrals at a scale defined by the average squark mass 
and neutralino  mass $Q^2\simeq (m_{\widetilde{q}}^2-m_{\widetilde Z_1}^2)$.
We employ the CTEQ3L parton distribution
function parametrization \cite{cteq} for numerical calculation. The most 
recent values of
the matrix element coefficients $f_{Tq}^{(N)}$ and $f_{TG}^{(N)}=1-\sum_{q}f_
{Tq}^{(N)}$ were compiled by \cite{jung} giving $f_{Tu}^{(p)}=0.019$, $f_{Td}^
{(p)}=0.041$, $f_{Tu}^{(n)}=0.023$, $f_{Td}^{(n)}=0.034$ and 
$f_{Ts}^{(p)}=f_{Ts}^{(n)}=0.14$.
We adopt the Saxon-Woods scalar form factor suggested in \cite{form}  
\begin{eqnarray*}
F(Q_r)=\frac{3j_1 (qR_0)}{qR_1}
e^{-\frac{1}{2}(qs)^2},
\end{eqnarray*}     
where $R_1=\sqrt{R^2-5s^2}$, $R=A^{\frac{1}{3}} \times 1.2\,\rm fm$, $j_1$ is a
spherical Bessel function and $s=1\,\rm fm$.

In the case of the spin interaction, the spin analogues of the parton
distibution functions are much less well known, and we take \cite{spindf} 
$\Delta u_p=\Delta d_n=0.78$, $\Delta d_p=\Delta u_n=-0.5$ and $\Delta s_p=
\Delta s_n=-0.16$. Theoretical predictions for the spin content of the two
nucleon groups in the nucleus and the spin form factor are very model
dependent. To make a consistent choice, we follow \cite{spinfrm}, where for 
$\rm ^{73}Ge$
$\langle S_p\rangle =0.03$, $\langle S_n\rangle =0.378$ and 
the form factor is given as
\begin{eqnarray*}
S(q)=(a_p+a_n)^2 S_{00}(q)+(a_p-a_n)^2 S_{11}(q)+(a_p^2-a_n^2) S_{01}(q)
\end{eqnarray*}     
and the $S_{ij}$ individual form factors are evaluated as polynomial fits to
data.

As a final step in the calculation, the differential rate $\frac{dR}{dQ_r}$
must be integrated over $Q_r$ ranging typically from $\sim 0$ to less than 
$100\,\rm keV$. We will show the rate $R$
in $\rm events/kg/day$ for a particular choice of the assumed local neutralino 
density $\rho_{\widetilde{Z_1}}=5\times 10^{-25}\, \rm g\, cm^{-3}$
\cite{flor}. The standard lore is that the local halo density
uncertainty should be roughly a factor of two; a more accurate prediction
based on galaxy formation models would be needed to reduce it.  

\section{Neutralino detection rates in mSUGRA parameter space}

We show our first numerical results for direct detection of 
neutralinos in Fig. 1, where we plot contours of scattering events/kg/day
for a $^{73}$Ge detector as a function of mSUGRA parameters $m_0\ vs.\ m_{1/2}$
with $A_0=0$, $\tan\beta =2$ and {\it a}) $\mu <0$ and {\it b}) $\mu >0$.
The region labelled TH is excluded by theoretical consideration:
either the LSP is charged or colored (not the lightest neutralino), or
radiative electroweak symmetry breaking is not properly attained. The region
labelled EX is excluded by collider searches for SUSY particles. By far the
strongest of these for mSUGRA is the recent limit from LEP2 
that $m_{\tw_1}>85$ GeV\cite{lep2lim}. We also show the contour of constant
$\Omega h^2=1$ (solid); beyond this contour, mSUGRA parameter space points
lead to universe with age less than 10 billion years. In addition, the 
dot-dashed contours correspond to $\Omega h^2=0.15$ and 0.4; the region
between these contours is favored by MDM cosmological models.

We find in frame {\it a}) that the direct dark matter detection rates are
uniformly low throughout the parameter space shown; for all points 
sampled, we found $R<0.01$/kg/day, which is less than 
the goal for such detectors at least
in the near-term future. In frame {\it b}), however, for $\mu >0$, we find 
larger rates for dark matter detection, with a significant fraction of
parameter space with $m_{1/2}<200$ GeV accessible to dark matter detectors
achieving a sensitivity of $R\agt 0.01$/kg/day. Note that the 
$R\agt 0.01$/kg/day region overlaps with the lower portion of the 
region favored by a MDM universe.

In Fig. 2, we show similar results, except now we take $\tan\beta =10$.
In Fig. 2, the region below the dotted contour is where $\Omega h^2<0.025$---
too small to account for the galactic rotation curves.
In frame {\it a}), we find considerably larger detection rates than for
Fig. 1{\it a}, with
$R$ reaching values $\sim 0.1$ in the lower-left. However, in this region,
$\Omega h^2<0.025$ so that the highest detection rates exist in an 
uninteresting region of relic density.
DM detectors with sensitivity $R\agt 0.01$/kg/day can probe just the
lower portion of the MDM cosmologically favored region.
Likewise, in frame {\it b}), 
event rates are larger as well than in Fig. 1{\it b}, 
with the entire region shown below
$m_{1/2}\simeq 200$ GeV accessible to DM detectors able to achieve a 
counting rate of $R=0.01$/kg/day.

Fig. 3 is similar in construction to Figs. 1 and 2, but with 
$\tan\beta =35$. Note, however, the expanded scale relative to Figs. 1 and 2.
The first peculiarity of note is that the TH region has expanded considerably.
This is due to the magnitude of the $\tau$ Yukawa coupling at 
large $\tan\beta$, which drives the $\ttau_1$ mass to lower values than
corresponding sleptons from the first two generations. The expanded TH 
region in the upper left is thus where $m_{\ttau_1}<m_{\tz}$, so that the
$\ttau_1$ is the LSP instead of the lightest neutralino.

In Fig. 3{\it a} and {\it b}, we see that the regions of cosmologically
interesting relic density $\Omega h^2$ are much larger than for the 
low $\tan\beta$ cases. 
In this case, the 
enhanced Higgs coupling to $b\bar{b}$ and $\tau\bar{\tau}$ 
at large $\tan\beta$ gives rise to 
a very broad resonance structure so that $s$-channel annihilation of
neutralinos is 
possible over a very large region of parameter space.
Note in particular that no upper limit on SUSY particle masses is evident
in these plots from the $\Omega h^2<1$ constraint.

We find in Fig. 3 that the dark matter detection rate $R$ 
has grown even larger for
low $m_{1/2}$ values, relative to Figs. 1 and 2, so that $R$ exceeds
1 event/kg/day in the lower left! In this region, however, $\Omega h^2<0.025$
so that again the largest detection rates are in a cosmologically
uninteresting region.
We see that $R$ can exceed values of 0.01/kg/day
for $m_{1/2}$ as high as 300 GeV, which corresponds to a reach in 
$m_{\tg}$ of $\sim 750$ GeV! 

The large increase in DM detection rate for large $\tan\beta$ has been noted
previously by Drees and Nojiri\cite{dn}. This is a fortuitous result: the
region of parameter space where DM detection is easiest is precisely 
the region of parameter space where collider detection of SUSY particles is
most difficult (this will be discussed in more detail in Sec. 4). We show the 
explicit variation in $R$ with $\tan\beta$ for the mSUGRA point
$m_0,m_{1/2},A_0=150,200,0$ GeV in Fig. 4. Here we see the event rate
increasing by more than one (two) orders of magnitude with $\tan\beta$ for
$\mu >0$ ($\mu <0$). Fig. 4 has some resemblance to Fig. 3 of the 
second paper of Ref. \cite{dn}; in our case (for a $^{73}$Ge detector 
instead of a spin-0 $^{76}$Ge detector), 
the large spin-spin interaction causes the main difference between the plots.

%In Fig. 5, we plot out some of the effective coupling constants 
%associated with the results from Fig. 4 (see Eq. A.6--A.8). The large effective
%coupling constants from scattering diagrams involving heavy Higgs bosons $H$
%accounts for the large increase in detection rates with $\tan\beta$.
%We also show the value of the spin coupling $a_p$ (see Eq. 2.8). 

For $^{73}$Ge,
the spin coupling is large, and can cause the axial-vector interaction
to exceed the scalar interaction even for nuclei as heavy as Ge. This is
shown in Fig. 5, where we plot the ratio $R_{spin}/R_{scalar}$ for the same
mSUGRA point as in Fig. 4. Here, we see that the spin interaction
actually dominates for $\tan\beta \sim 6$, for $\mu <0$. This is 
contrary to naive expectations that the spin interaction is always
sub-dominant for nuclei with $A\agt 20$.

In Fig. 6, we show an example of the variation in dark matter detection
rate versus variation in the trilinear soft-breaking term $A_0$, for
$m_0,m_{1/2}=150,200$ GeV, and $\tan\beta =10$. From this plot, we see 
that the DM detection rate can vary with $A_0$ by an order of magnitude; 
this sort of variation must be taken into account if dark matter detection
experiments ever try to obtain limits on mSUGRA parameter space.

Finally, we note that a seasonal variation in the DM detection rate is 
expected\cite{jung}. This is caused by accounting for the earth's velocity
about the sun, while at the same time accounting for the sun's 
velocity about the galactic center. In Fig. 7, we plot the DM detection
rate as a function of month of the year (beginning with Jan. 1). The 
maximum detection rate occurs around June 2, although the seasonal
variation amounts to less than a per cent, so that very high counting rates 
would be necessary to detect this. We note as well that if DM detectors
are sensitive to the direction of collision products, these should also
depend on the time of day and season of the year.

\section{Comparison with SUSY searches at colliders}

In this section, we compare our results for direct detection of neutralino DM
with constraints from $B\to X_s \gamma$ searches, and with expectations
for various collider searches. In these comparisons, we are restricted to
values of $\tan\beta <10$, for which detailed collider and $B\to X_s \gamma$
calculations are available. Only recently has the event generator ISAJET
been upgraded to handle large $\tan\beta$ cases, so detailed studies
for collider expectations still need to be made\cite{ltanbl}.

The rare decay $B\to X_s \gamma$ has been 
shown to yield rather strong constraints on supersymmetric models, due to loop
amplitudes containing charginos, neutralino, gluinos, squarks and charged
Higgs bosons\cite{bsgpapers}. 
In a recent paper\cite{bsg}, a QCD-improved calculation
of the $B\to X_s \gamma$ branching fraction has been made for the mSUGRA model.
The sensitivity to variations in the QCD renormalization scale has been reduced
considerably compared to previous results. 
In this paper, only chargino, charged Higgs and $W$-boson loops have been 
included, which is appropriate for small to moderate values of $\tan\beta$.
Comparison of calculated 
branching fractions to recent results from the CLEO experiment has resulted
in identification of regions of mSUGRA model parameter space which are
excluded at 95\% CL.
For large $\tan\beta$, $\tg$ and $\tz_i$ loops will also be 
relevant (see Borzumati, Drees and Nojiri, Ref. \cite{bsgpapers}).
In this case, the derived constraints will depend sensitively upon details of 
the assumed structure of high scale squark mass matrices, so that the 
implications will be much more model dependent. 
%This issue is undergoing further examination.

In Fig. 9, we again show the $m_0\ vs.\ m_{1/2}$ plane  for $\tan\beta =2$.
This time, we show in Fig. 9{\it a} the region excluded by
the 95\% CL CLEO result compared to mSUGRA model calculations. We exclude 
parameter space points where the  $B\to X_s \gamma$ branching ratio 
falls outside the CLEO 95\% CL limits for {\it all} choices of 
renormalization scale $\frac{m_b}{2}<Q<2m_b$.
The excluded region is in the lower left of frame {\it a}), where the
dark matter detection rate is largest, although still smaller than
$R=0.01$/kg/day. In this frame, we also show the region that can be searched by
the LEP2 $e^+e^-$ collider operating at $\sqrt{s}=190$ GeV, and
accumulating $\sim 500$ pb$^{-1}$ of integrated luminosity (dotted curve).
The left-side bulge in this contour is where LEP2 is sensitive to 
selectron searches, while below the right-hand side, 
which asymptotically approaches $m_{1/2}\simeq 100$, is where LEP2 is sensitive
to chargino pair searches\cite{bbmt}. We also note that LEP2 is sensitive 
to almost the entire plane shown via the $e^+e^-\to Zh$ search channel; 
in this case, however, it will be difficult to tell the light SUSY Higgs $h$
from the SM Higgs boson. Finally, we also show the dashed contour, 
which is the reach of the Tevatron Main Injector (MI: $\sqrt{s}=2$ TeV;
integrated luminosity $=2$ fb$^{-1}$). This latter curve mainly results 
from regions where the clean trilepton signal from $\tw_1\tz_2\to 3\ell$ is 
observable above SM backgrounds\cite{tevreach}. By comparing all the contours
of Fig. 9{\it a}, we note that the low $m_{1/2}$ region of the MDM-favored
region is excluded by $B\to X_s \gamma$, which also excludes much of the region
open to SUSY discovery by LEP2 and Tevatron MI searches. If mSUGRA is correct,
with $\tan\beta =2$ and $\mu <0$, then LEP2 may well discover the light 
Higgs scalar $h$, but probably there will be no direct detection of SUSY DM
and SUSY discovery will have to wait for the CERN LHC
$pp$ collider, which has the ability to explore the entire SUSY parameter
space with relatively low luminosity\cite{lhc}.

In Fig. 9{\it b}, none of the $m_0\ vs.\ m_{1/2}$ plane is excluded by
$B\to X_s \gamma$. The reach of LEP2 is almost always
below the MDM-favored region. However, both the Tevatron MI as well as
DM detectors sensitive to $R\simeq 0.01$/kg/day can explore the lower
limits of the MDM region. In this case, the MI has somewhat of a 
better reach for SUSY than direct DM detection experiments, while LEP2 would stand
again a good chance to find the light Higgs scalar $h$\cite{bbmt}.

In Fig. 10{\it a}, we show the plane for $\tan\beta =10$ and $\mu <0$. The 
outstanding feature here is that $B\to X_s \gamma$ data exclude almost the
whole plane below $m_{1/2}\sim 350$ GeV 
(corresponding to $m_{\tg}\alt 900$ GeV), including essentially all of the
MDM-favored region\cite{bsg}! If mSUGRA exists in this parameter plane, then
direct DM detection experiments, LEP2 and Tevatron MI will find no evidence
of SUSY, and SUSY discovery will have to await the LHC (although 
$B\to X_s \gamma$ experiments may have a strong hint of new physics).

Fig. 10{\it b} shows the $m_0\ vs.\ m_{1/2}$ plane for $\tan\beta =10$ and 
$\mu >0$. In this case, none of the plane shown is excluded by 
$B\to X_s \gamma$. In fact, the mSUGRA region around $m_{1/2}\simeq 200$ GeV 
gives 
a better match to CLEO data than does the SM! For this case, the reach of LEP2
is below the MDM region, and in addition the Higgs scalar $h$ is too heavy for
detection at LEP2. The Tevatron MI can explore a portion of the MDM 
favored region. However, a DM detector with sensitivity $R\simeq 0.01$/kg/day
will be sensitive to the additional region with $m_{1/2}\alt 200$ GeV, and
$m_0\agt 150$ GeV, which is inaccessible to Tevatron or LEP2 searches. This 
region includes the entire MDM-favored $s$-channel annihilation corridor
for which $m_{1/2}\sim 180$ GeV. This latter region is also favored by
the $B\to X_s \gamma$ measurement\cite{bsg}.

Finally, we coment upon the large $\tan\beta$ region, for which detailed
collider studies and $B\to X_s\gamma$ calculations have yet to be made. 
Drees and Nojiri pointed out that
as $\tan\beta$ increase, the Higgs
pseudoscalar mass $m_A$ decreases significantly, so that over much of 
parameter space the neutralino relic density decreases, mainly due to
$s$-channel $\tz\tz\to A,H\to b\bar{b}$ annihilation reactions\cite{relic}.
We have verified this with the relic density contours presented in Fig. 3.
In a recent paper\cite{ltanbl}, detailed calculations of sparticle masses,
production and decay processes at large $\tan\beta$ were reported. It was noted
that for large $\tan\beta$, the $\tw_1$ and $\tz_2$ branching ratios to
$\tau$ leptons and $b$-quarks increases due to Yukawa coupling effects, which
leads to a diminution of the corresponding branchings to easily detectable
$e$ and $\mu$ states. This generally ought to make SUSY detection much more
difficult for the Tevatron MI than corresponding mSUGRA points with low
$\tan\beta$. However, we note from Fig. 3 that the large $\tan\beta$ region
is precisely where DM detection rates can be largest. Hence, if $\tan\beta$ is
large, it is possible that the first evidence for SUSY might come from
direct dark matter detection, rather than from LEP2 or MI searches. It is 
expected that the CERN LHC $pp$ collider will still be able to cover the
entire mSUGRA parameter space even if $\tan\beta$ is large, at least via
multijet$+\eslt$ searches\cite{ltanbl}.

\section{Summary and conclusions}

In this paper, we have presented expected event rates for direct detection
of neutralinos by cryogenic dark matter detectors. To be specific, we chose
to perform calculations for a $^{73}$Ge detector. Many other choices of 
materials are possible, and analogous calculations can be made by using 
different nuclear $A$, $Z$ and $J$ values, and by using different nuclear form
factors. Our main results were presented in Figs. 1--3 
as functions of mSUGRA parameter space. When interpreting these results, the
intrinsic uncertainties in the calculations should be kept in mind. 
Roughly, we expect a factor of $\sim 2$ uncertainty from 
each of {\it i.}) calculations in leading-log QCD, {\it ii.}) uncertainty 
in knowledge of the {\it local} dark matter relic density and {\it iii.}) 
uncertainty 
in nuclear form factors and in quark contributions to nucleon spin.
Simplistically adding these in quadature would imply our results for event 
rates are only reliable to a factor 3--4. In addition, variation of 
the parameter $A_0$ can cause changes in the DM detection rates by up to
an order of magnitude (see Fig. 6).

One of our main goals in this paper was to compare the reach for 
supersymmetry by dark matter detectors against the reach for SUSY via collider 
experiments and rare decay searches. A dark matter detector reaching a 
sensitivity of $R\sim 0.01$/kg/day usually would have a better reach in
mSUGRA parameter space than LEP2 would for SUSY particles (but not if one
includes LEP2 sensitivity to Higgs bosons). Dark matter detectors can 
be comparable to the Tevatron MI in terms of reach for mSUGRA 
for low values of $\tan\beta$.
However, for large $\tan\beta$, dark matter detectors have an {\it increased}
event rate, whereas the reach of the Tevatron MI will likely diminish 
relative to capabilities at modest values of $\tan\beta$. 
For the large $\tan\beta$ case, the first direct 
evidence for SUSY may well come from direct detection experiments.
We note that even if dark matter detectors only achieve a 
sensitivity of 0.1/kg/day, they would still have a substantial reach for SUSY
in the large $\tan\beta$ region. 
The relative capabilities 
at low and high $\tan\beta$ underscores another facet of complementarity
between collider search experiments, and direct detection of dark matter.
Obviously, if one detects SUSY at collider experiments, it would still be 
fascinating to verify the existence and properties of neutralino dark
matter.

%%%%%%%%%%%%%%%%%%%%%%%%% ACKNOWLEDGEMENTS %%%%%%%%%%%%%%%%%%%%%%%%%%%%%%%%%%%%%

%\newpage
\acknowledgments

We thank M. Drees and X. Tata for comments on the manuscript.
This research was supported in part by the U.~S. Department of Energy
under grant number DE-FG-05-87ER40319.

\newpage
\appendix
\section{Effective Couplings for Neutralino-Nucleon Scattering}

\def\theequation{A.\arabic{equation}}
\setcounter{equation}{0}

In this Appendix we summarize effective couplings entering the scalar and 
spin Lagrangian for neutralino scattering on nucleons, and express them in 
terms of the MSSM coupling constants and masses. Most of the time we follow 
reference \cite {dn}. 

Both interactions are mediated by squarks, Higgs bosons or the $\rm Z^0$ boson
connecting either directly or through a loop the neutralino lines of 
propagation with quark (gluon) lines in the nucleon. The lightest
neutralino-quark-squark Lagrangian reads
\begin{eqnarray}
{\cal L}_{\widetilde{Z}_{\ell} q \widetilde{q}}=\sum_{i=1,2} \bar{\widetilde 
{Z}_{\ell}} (a_{\widetilde{q}_i}+b_{\widetilde{q}_i}\gamma_5 )q \widetilde{q}_i
+ h.c.  
\end{eqnarray}
and the $a$ and $b$ coefficients take the form
\begin{eqnarray}    
a_{\widetilde{q}_i}&=&\frac{1}{\sqrt{2}} \biggl\{ -M_{i1} [e e_q N_{\ell 1}^{'}
+\frac{g}{\cos \theta_W} N_{\ell 2}^{'} (T_{3q}-e_q \sin^{2} \theta_W )
+\frac{g m_q}{2 m_W} N_{\ell x} ] \biggr. \nonumber \\
&+& \biggl. M_{i2} [e e_q N_{\ell 1}^{'}
-\frac{g}{\cos \theta_W} N_{\ell 2}^{'} e_q \sin^{2} \theta_W
-\frac{g m_q}{2 m_W} N_{\ell x} ] \biggr\} ,   \\
b_{\widetilde{q}_i}&=&\frac{1}{\sqrt{2}} \biggl\{ -M_{i1} [e e_q N_{\ell 1}^{'}
+\frac{g}{\cos \theta_W} N_{\ell 2}^{'} (T_{3q}-e_q \sin^{2} \theta_W )
-\frac{g m_q}{2 m_W} N_{\ell x} ] \biggr. \nonumber \\
&+& \biggl. M_{i2} [-e e_q N_{\ell 1}^{'}
+\frac{g}{\cos \theta_W} N_{\ell 2}^{'} e_q \sin^{2} \theta_W
-\frac{g m_q}{2 m_W} N_{\ell x} ] \biggr\} ,    \
\end{eqnarray}
where 
\begin{eqnarray}    
\pmatrix{\widetilde{q}_1\cr \widetilde{q}_2}=\pmatrix{M_{11}&M_{12}\cr
M_{21}& M_{22}} , \pmatrix{\widetilde{q}_L \cr \widetilde{q}_R}=\pmatrix{\cos \theta_q & \sin\theta_q \cr
-\sin\theta_q & \cos \theta_q} \pmatrix{\widetilde{q}_L \cr \widetilde{q}_R} ,
\end{eqnarray}
and $x=3(4)$ for a down (up) type quark. The $4\times 4$ $N$ matrix diagonalizes
the neutralino mass matrix and
\begin{eqnarray}    
\pmatrix{N'_{j1}\cr N'_{j2}\cr}=\pmatrix{\cos \theta_W & \sin\theta_W \cr
-\sin\theta_W & \cos \theta_W} \pmatrix{N_{j1} \cr N_{j2}}. 
\end{eqnarray}
The $f_q$ coupling in (2.15) can be split into two parts 
\begin{eqnarray}
f_q=f_q^{(\widetilde{q})}+f_q^{(H)} , 
\end{eqnarray}
where the squark part is
\begin{eqnarray}
f_q^{(\widetilde{q})}=-\frac{1}{4}\sum_{i=1,2} 
\frac{a_{\widetilde{q}_i}^2-b_{\widetilde{q}_i}^2}{m_{\widetilde{q_i}}^2 
-(m_{\widetilde{Z_{1}}}+m_q)^2} , 
\end{eqnarray}
and the Higgs exchange part is  
\begin{eqnarray}
f_q^{(H)}=m_q \sum_{j=1,2} \frac{c_{\widetilde{Z}}^{(j)} c_{q}^{(j)}}
{m_{H_j}^2}. 
\end{eqnarray}
Mixing in the Higgs sector results in 
\begin{eqnarray}
c_{\widetilde{Z}}^{(1)}=\frac{1}{2}(g N_{\ell 2}-g' N_{\ell 1})
(N_{\ell 3}\sin \alpha+N_{\ell 4}\cos\alpha)
\end{eqnarray}
for the lighter CP-even Higgs and 
\begin{eqnarray}
c_{\widetilde{Z}}^{(2)}=\frac{1}{2}(g N_{\ell 2}-g' N_{\ell 1})
(N_{\ell 4}\sin \alpha-N_{\ell 3}\cos\alpha)
\end{eqnarray}
for the heavier Higgs, where $\alpha$ is the Higgs mixing angle. 
The quark coefficients are evaluated as 
\begin{eqnarray}
c_{q}^{(i)}=\frac{g}{2 m_W} r_q^{(i)}
\end{eqnarray}
with 
\begin{eqnarray}
r_{u}^{(1)}=-\frac{\sin\alpha}{\sin\beta},\ \ \ r_{u}^{(2)}=-\frac{\cos\alpha}{\sin\beta}
\end{eqnarray}
for the up type quarks and 
\begin{eqnarray}
r_{d}^{(1)}=-\frac{\cos\alpha}{\cos\beta},\ \ \ r_{d}^{(2)}=\;\; \frac{\sin\alpha}{\cos\beta}
\end{eqnarray}
for the down type quarks.
The quark tensor contribution coupling in (2.15) can be expressed as
\begin{eqnarray}
g_q=-\frac{1}{4}\sum_{i=1,2} 
\frac{a_{\widetilde{q}_i}^2+b_{\widetilde{q}_i}^2}{[ m_{\widetilde{q_i}}^2 
-(m_{\widetilde{Z_{1}}}+m_q)^2]^2} . 
\end{eqnarray}
The gluon part of the scalar effective Lagrangian is fully determined by
\begin{eqnarray}
b=-T_{\widetilde{q}}+B_D+B_S-\frac{m_{\widetilde{Z_{1}}}}{2} B_{2S}
-\frac{m_{\widetilde{Z_{1}}}^2}{4} (B_{1D}+B_{1S}),
\end{eqnarray}
where 
\begin{eqnarray}
T_{\widetilde{q}}&=&\frac{1}{96\pi}\sum_{j=1,2} \frac{c_{\widetilde{Z}}^{(j)}}
{m_{H_j}^2} \sum_{q,i} \frac{c_{\widetilde{q_i}}^{(j)}}
{m_{\widetilde{q_i}}^2} \\ 
B_D&=&\frac{1}{32\pi} \sum_{q,i} (a_{\widetilde{q}_i}^2-b_{\widetilde{q}_i}^2)
m_q I_1(m_{\widetilde{q_i}},m_q,m_{\widetilde{Z_{1}}})\\
B_S&=&\frac{1}{32\pi} \sum_{q,i} (a_{\widetilde{q}_i}^2+b_{\widetilde{q}_i}^2)
m_{\widetilde{Z_{1}}} I_2(m_{\widetilde{q_i}},m_q,m_{\widetilde{Z_{1}}})\\
B_{1D}&=&\frac{1}{12\pi} \sum_{q,i} (a_{\widetilde{q}_i}^2-b_{\widetilde{q}_i}^2)
m_q I_3(m_{\widetilde{q_i}},m_q,m_{\widetilde{Z_{1}}})\\
B_{1S}&=&\frac{1}{12\pi} \sum_{q,i} (a_{\widetilde{q}_i}^2+b_{\widetilde{q}_i}^2)
m_{\widetilde{Z_{1}}} I_4(m_{\widetilde{q_i}},m_q,m_{\widetilde{Z_{1}}})\\
B_{2S}&=&\frac{1}{48\pi} \sum_{q,i} (a_{\widetilde{q}_i}^2+b_{\widetilde{q}_i}^2)
I_5(m_{\widetilde{q_i}},m_q,m_{\widetilde{Z_{1}}}).\
\end{eqnarray}
The squark effective couplings in (B.16) are
\begin{eqnarray}
c_{\widetilde{q_1}}^{(j)}&=&\frac{g m_Z}{\cos\theta_W} s^{(j)} (T_{3q}\cos^2\theta_q
-e_q\sin^2\theta_W \cos 2\theta_q) \nonumber \\
&+&\frac{g m_q^2}{m_W} r_q^{(j)}-\frac{g m_q \sin 2\theta_q}{2 m_W} (\mu {r'}_q^{(j)}
-A_q r_q^{(j)}) ,\\
c_{\widetilde{q_2}}^{(j)}&=&\frac{g m_Z}{\cos\theta_W} s^{(j)} (T_{3q}\sin^2\theta_q
+e_q\sin^2\theta_W \cos 2\theta_q) \nonumber \\
&+&\frac{g m_q^2}{m_W} r_q^{(j)}+\frac{g m_q \sin 2\theta_q}{2 m_W} (\mu {r'}_q^{(j)}
-A_q r_q^{(j)}) ,\
\end{eqnarray}
where
\begin{eqnarray}
s^{(1)}=-\cos (\alpha+\beta)&&
s^{(2)}= \sin (\alpha+\beta)
\end{eqnarray}
and 
\begin{eqnarray}
{r'}_{u}^{(1)}=-\frac{\cos\alpha}{\sin\beta},\ \ \
{r'}_{u}^{(2)}=\frac{\sin\alpha}{\sin\beta}, \\
{r'}_{d}^{(1)}=-\frac{\sin\alpha}{\cos\beta},\ \ \
{r'}_{d}^{(2)}=-\frac{\cos\alpha}{\cos\beta}.
\end{eqnarray}                              
Loop integrals $I_1$ - $I_5$ are given by Eqs. (B1a-e) in Ref. \cite{dn}
(one must take care to correct the typo noted in Ref. \cite{jung}).

For the only effective coupling needed in the spin dependent Lagrangian we have
\begin{eqnarray}
d_q=\frac{1}{4}\sum_{i=1,2} 
\frac{a_{\widetilde{q}_i}+b_{\widetilde{q}_i}}{m_{\widetilde{q_i}}^2 
-(m_{\widetilde{Z_{1}}}+m_q)^2}-\frac{g^2}{4 m_W^2} {O''}^R T_{3q},
\end{eqnarray}
where
\begin{eqnarray}
{O''}^R=\frac{1}{2} (N_{\ell 4}^{2}-N_{\ell 3}^{2})
\end{eqnarray}
is determined by the neutralino mass matrix diagonalizing matrix $N$.

%%%%%%%%%%%%%%%%%%%%% REFERENCES %%%%%%%%%%%%%%%%%%%%%%%%%%%%%%%%%%%%%%%%%%%%%%
%

%
\newpage
%
%%%%%%%%%%%%%%%%%%%%%%%%%% TABLES %%%%%%%%%%%%%%%%%%%%%%%%%%%%%%%%%%%%%%%%%%%
%

%%%%%%%%%%%%%%%%%%%%%% FIGURE CAPTIONS %%%%%%%%%%%%%%%%%%%%%%%%%%%%%%%%%%%%%%
\begin{figure}
\iftightenlines\epsfxsize=5in
\centerline{\epsfbox{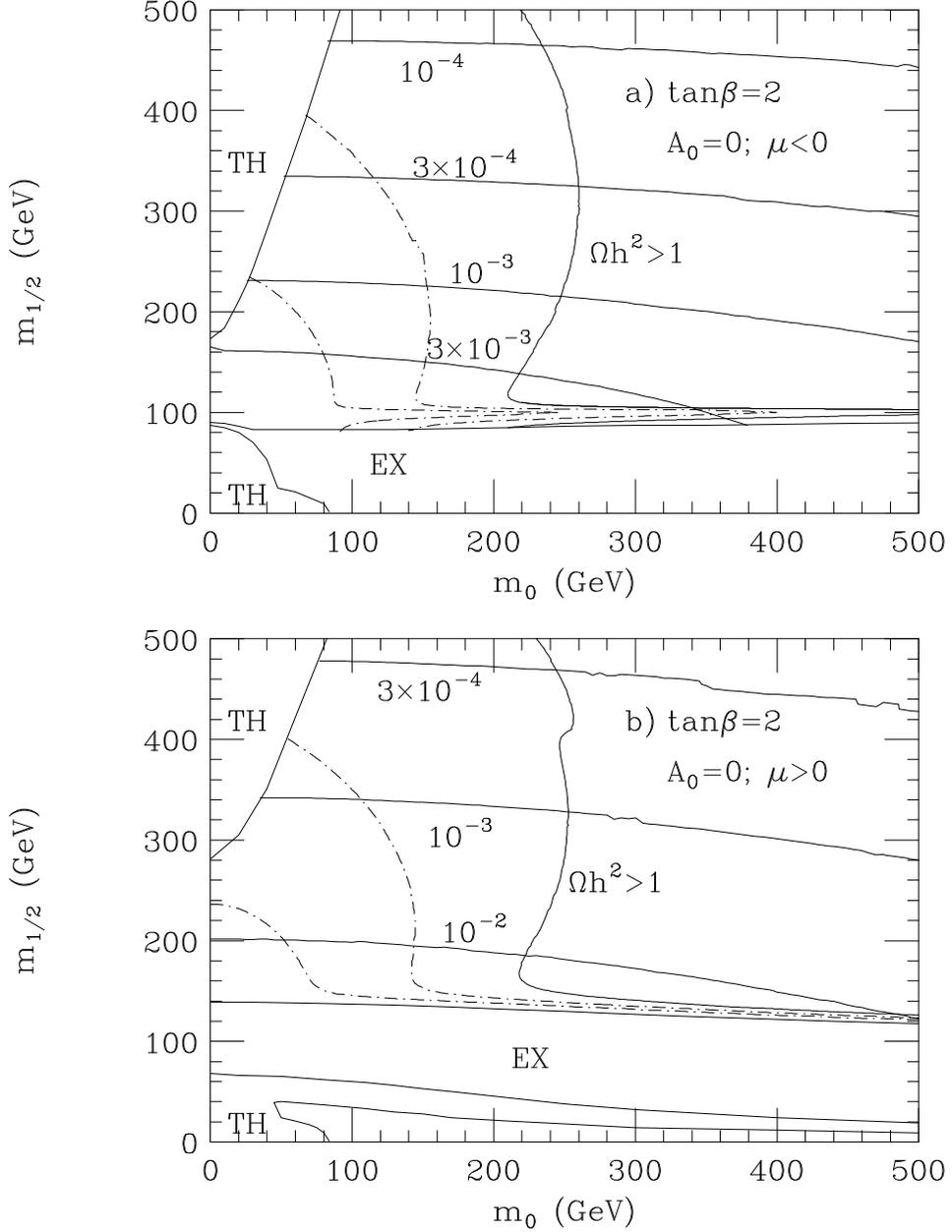}}
\medskip\fi
\caption[]{A plot of contours of neutralino scattering events/kg/day in a 
$^{73}$Ge detector, for mSUGRA parameters $A_0=0$, $\tan\beta =2$ and
{\it a}) $\mu <0$ and {\it b}) $\mu >0$, in the $m_0\ vs.\ m_{1/2}$ plane.
The regions labelled TH are excluded by theoretical considerations,
while the EX regions are excluded by collider searches for SUSY 
particles. The region to the right of the solid contour is excluded by
$\Omega h^2>1$. We also show contours of neutralino relic density
$\Omega h^2=0.15$ and $0.4$; the region in between is favored by models 
of a MDM universe.
}
\label{FIG1}
\end{figure}
\begin{figure}
\iftightenlines\epsfxsize=5in
\centerline{\epsfbox{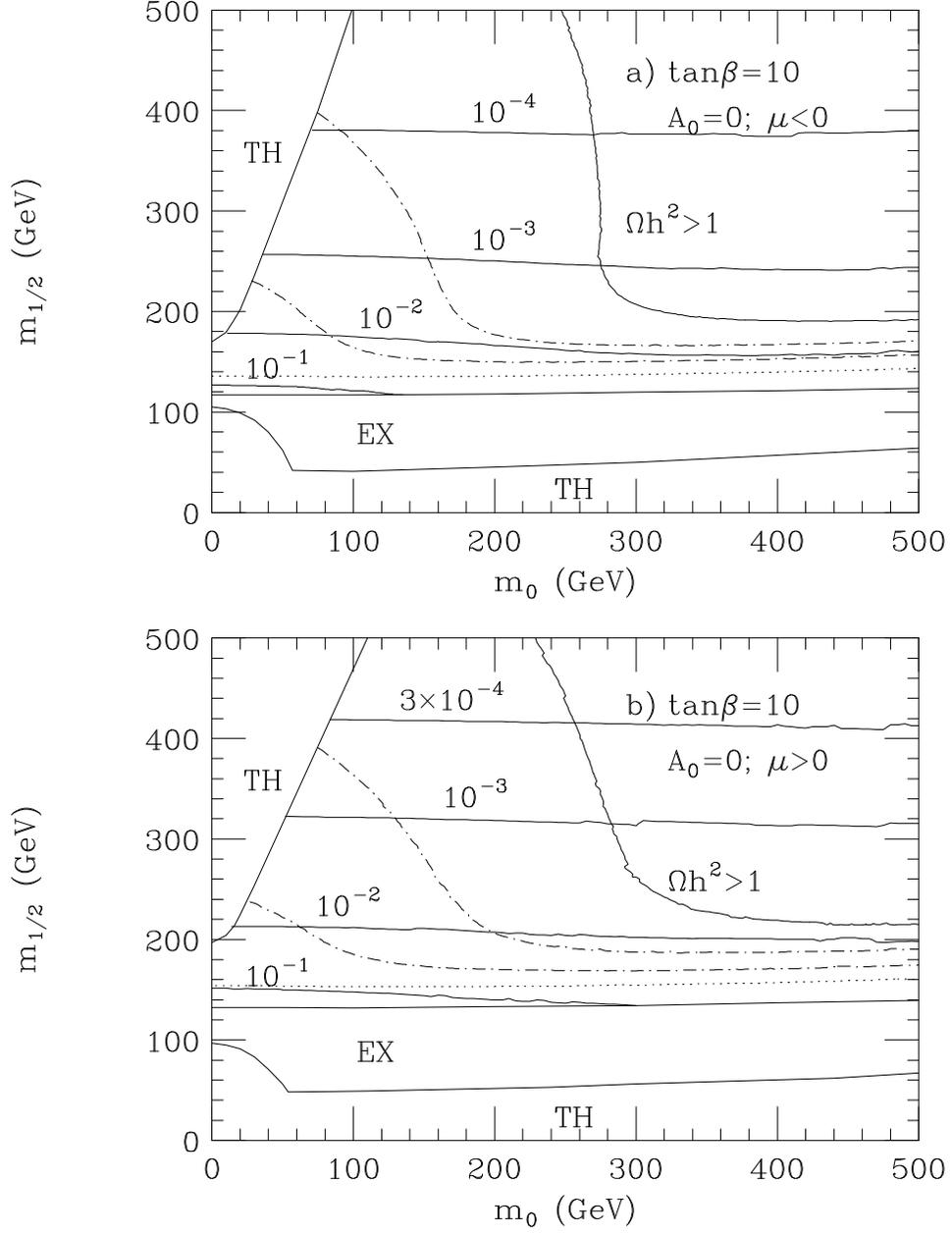}}
\medskip\fi
\caption[]{Same as Fig. 1, except for $\tan\beta =10$. Below the dotted
contour is where $\Omega h^2<0.025$. 
}
\label{FIG2}
\end{figure}
\begin{figure}
\iftightenlines\epsfxsize=5in
\centerline{\epsfbox{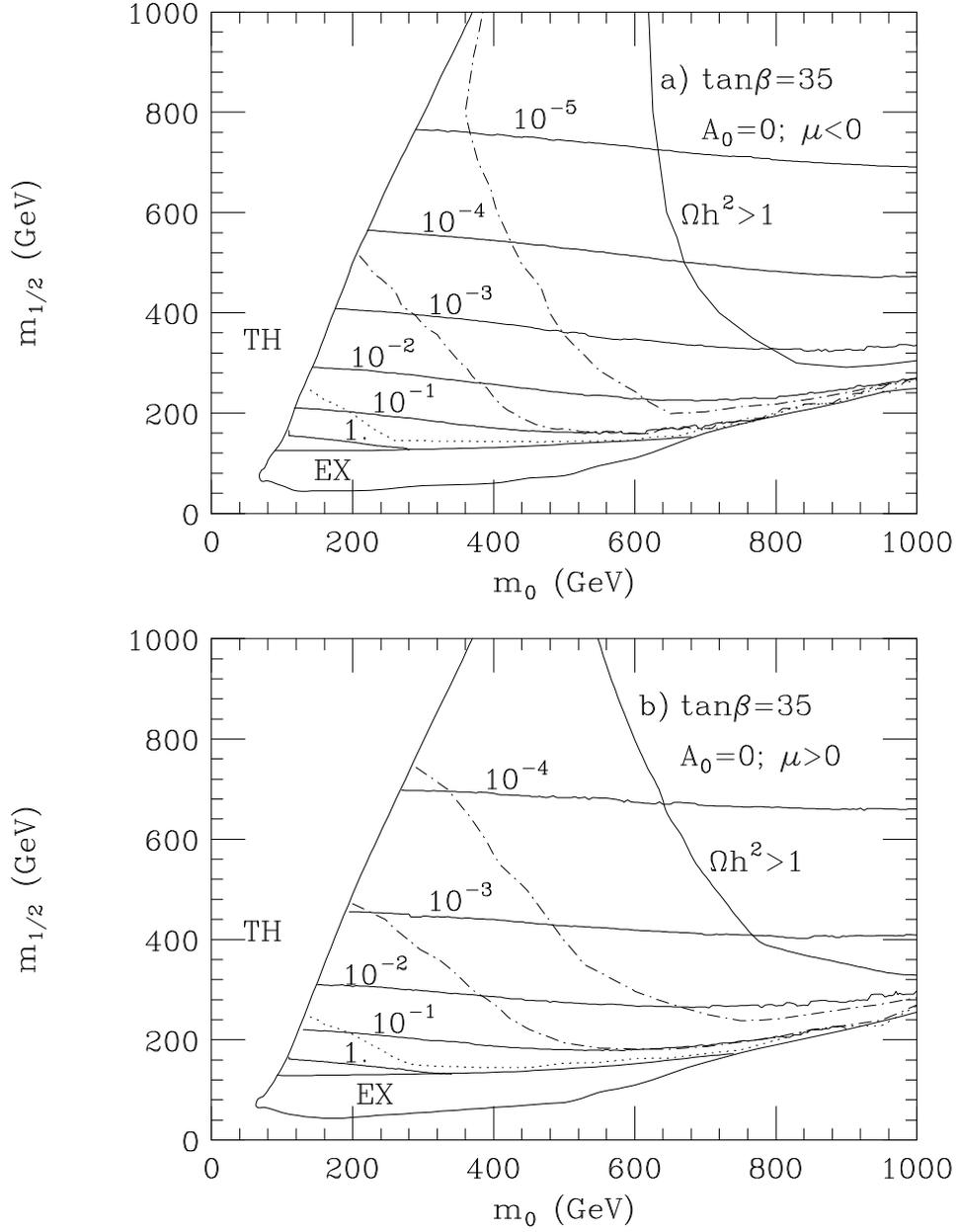}}
\medskip\fi
\caption[]{Same as Fig. 1, except for $\tan\beta =35$. Note, however, the
expanded scale relative to Fig. 1 and 2.
}
\label{FIG3}
\end{figure}
\begin{figure}
\iftightenlines\epsfxsize=6in
\centerline{\epsfbox{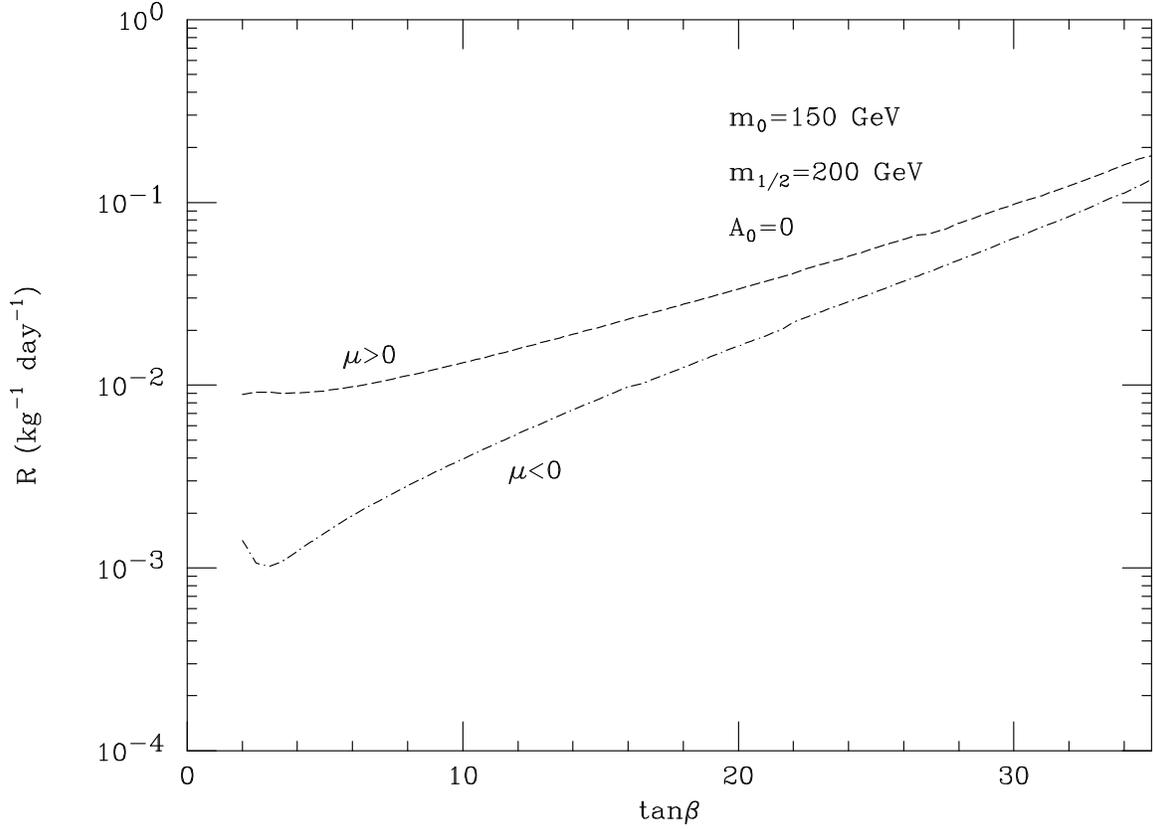}}
\medskip\fi
\caption[]{A plot of neutralino scattering events/kg/day in a 
$^{73}$Ge detector, for mSUGRA parameters 
$m_0 ,m_{1/2} ,A_0 =$150, 200, and 0 GeV versus $\tan\beta$ for $\mu <0$
and $\mu >0$.
}
\label{FIG4}
\end{figure}
%
%\begin{figure}
%\caption[]{A plot of effective coupling constants for the mSUGRA point
%$m_0, m_{1/2}, A_0=$150, 200, and 0 GeV versus 
%$\tan\beta$ for {\it a}) $\mu <0$
%and {\it b}) $\mu >0$. The coupling $a_p$ enters the spin-spin
%interaction, while the remaining curves are for the scalar coupling
%contribution to the detection rate.
%}
%\end{figure}
%
\begin{figure}
\iftightenlines\epsfxsize=6in
\centerline{\epsfbox{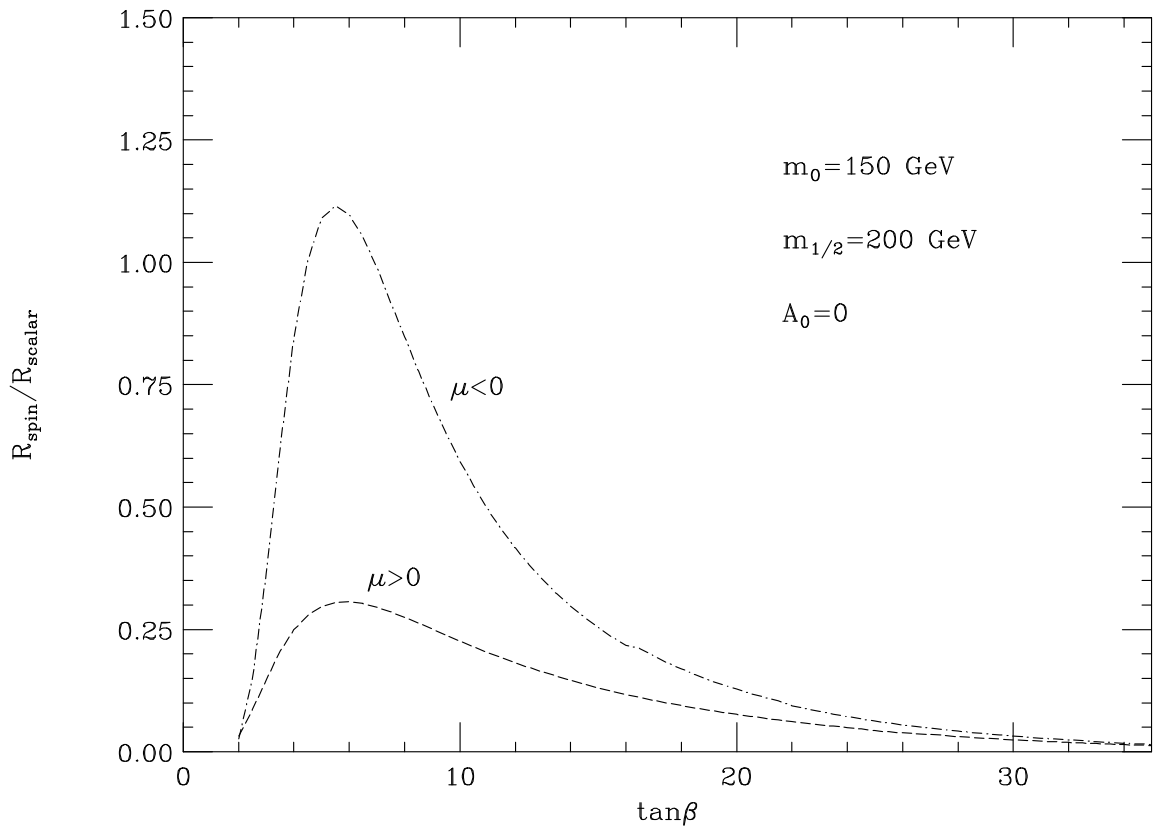}}
\medskip\fi
\caption[]{A plot of the ratio of neutralino scattering events from 
the spin-spin interaction over that from scalar interactions
for mSUGRA parameters $m_0 ,m_{1/2} ,A_0 =$150, 200, and 0 GeV,
 versus $\tan\beta$ for $\mu <0$ and $\mu >0$.
}
\label{FIG5}
\end{figure}
\begin{figure}
\iftightenlines\epsfxsize=6in
\centerline{\epsfbox{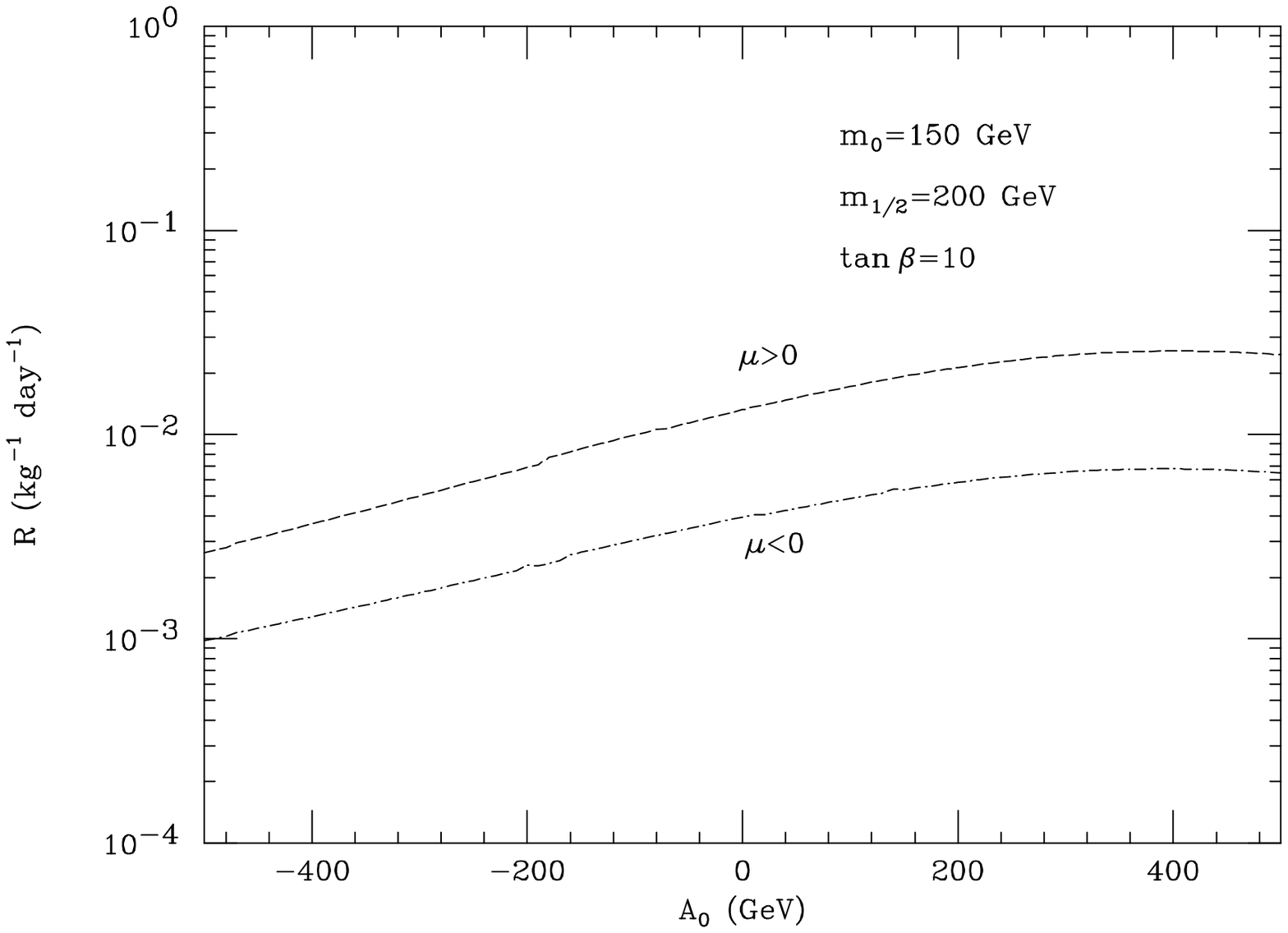}}
\medskip\fi
\caption[]{A plot of neutralino scattering events/kg/day in a 
$^{73}$Ge detector, for mSUGRA parameters 
$m_0 ,m_{1/2} =$150, 200 GeV with $\tan\beta =10$, versus the 
parameter $A_0$,  for $\mu <0$ and $\mu >0$.
}
\label{FIG6}
\end{figure}
\begin{figure}
\iftightenlines\epsfxsize=6in
\centerline{\epsfbox{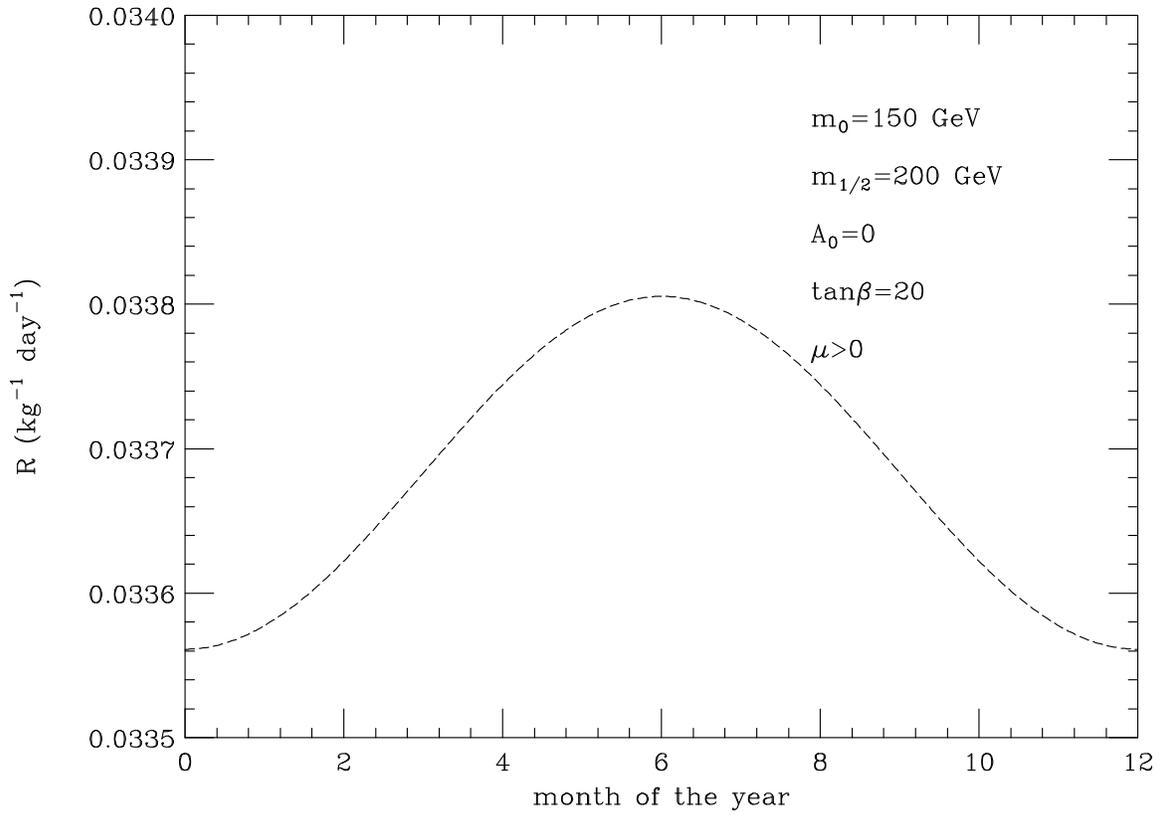}}
\medskip\fi
\caption[]{A plot of neutralino scattering events/kg/day in a 
$^{73}$Ge detector for mSUGRA parameters 
$m_0 ,m_{1/2}, A_0 =$150, 200, 0 GeV with $\tan\beta =20$ and $\mu >0$,
versus month of the year, from Jan. 1 to Dec. 31, showing the 
expected seasonal variation in dark matter detection rate.
}
\label{FIG7}
\end{figure}
\begin{figure}
\iftightenlines\epsfxsize=5in
\centerline{\epsfbox{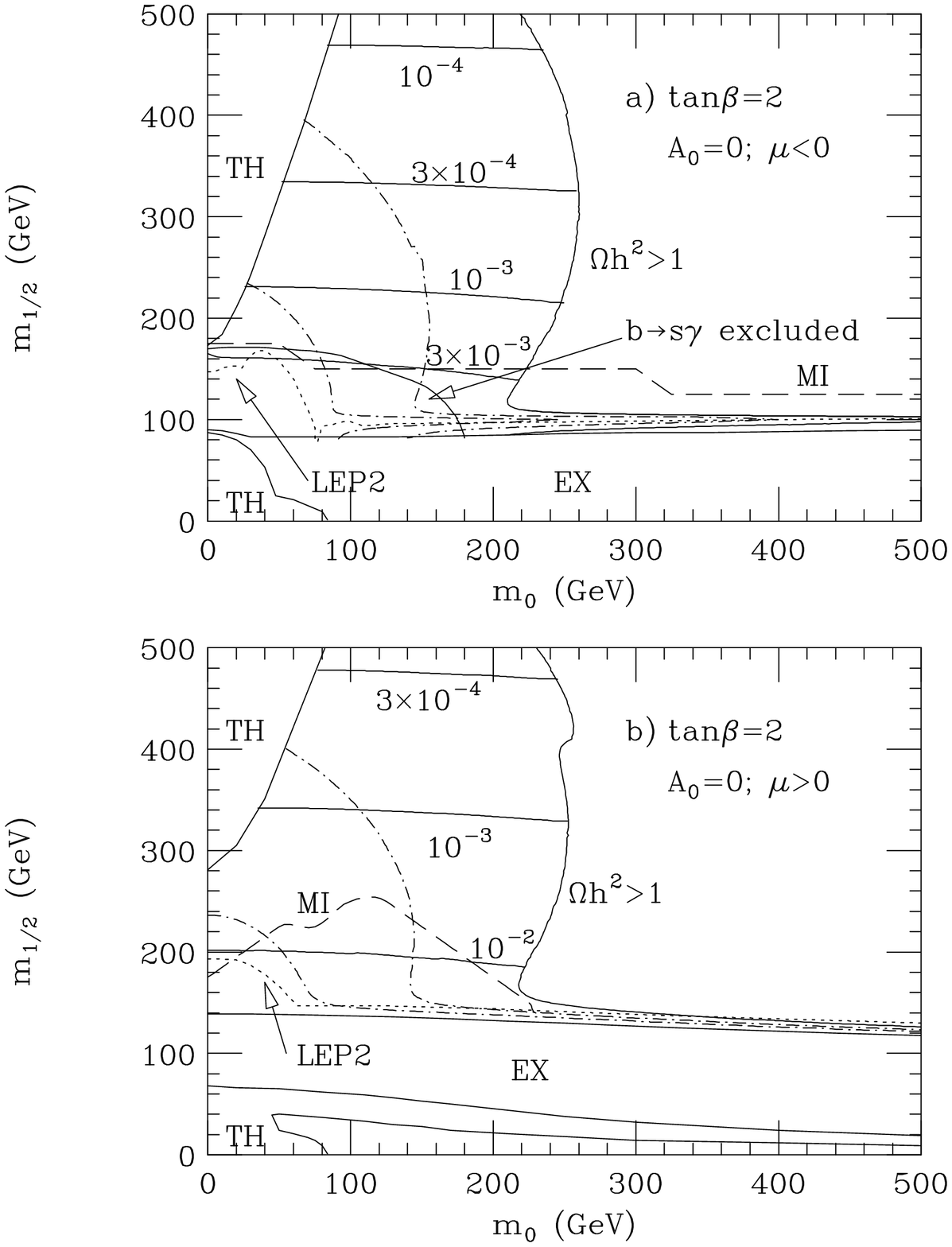}}
\medskip\fi
\caption[]{Same as Fig. 1, but with in addition $b\to s\gamma$ exclusion
contours, plus the reach contours for LEP2 and Fermilab Tevatron MI 
experiments for detecting SUSY.
}
\label{FIG8}
\end{figure}
\begin{figure}
\iftightenlines\epsfxsize=5in
\centerline{\epsfbox{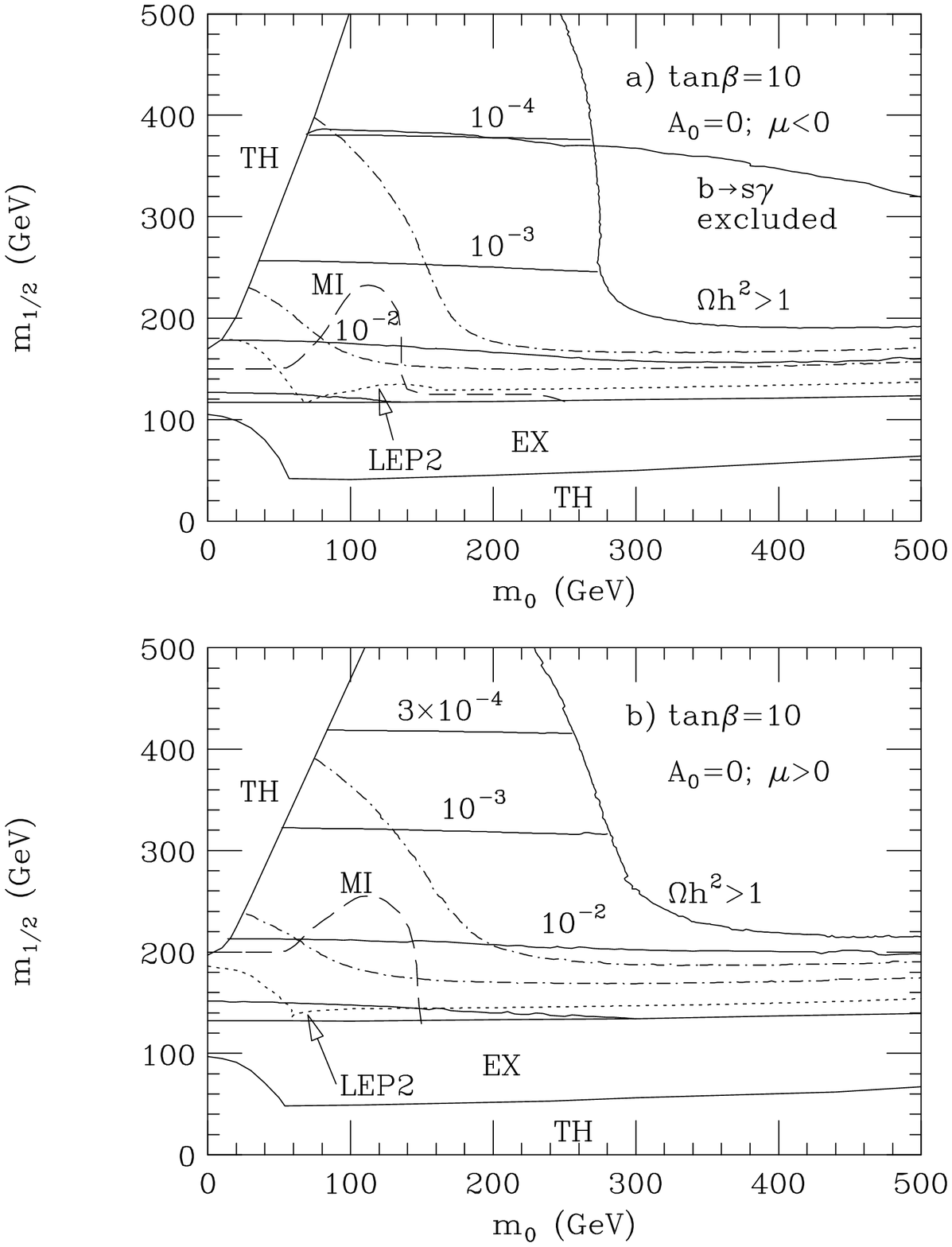}}
\medskip\fi
\caption[]{Same as Fig. 2, but with in addition $b\to s\gamma$ exclusion
contours, plus the reach contours for LEP2 and Fermilab Tevatron MI 
experiments for detecting SUSY.
}
\label{FIG9}
\end{figure}

\end{document}